\documentclass[journal=jacsat,manuscript=article]{achemso}

\usepackage[version=3]{mhchem} 

\usepackage{graphicx}
\usepackage{dcolumn}
\usepackage{bm}

\usepackage[utf8]{inputenc}
\usepackage{etoolbox}
\usepackage{xcolor}
\usepackage{hyperref}

\author{Ashesh Ghosh}%
 \email{asheshg@stanford.edu}
 \affiliation{$^{1}$Department of Chemical Engineering, Stanford University, Stanford CA 94305 }

\title[An \textsf{achemso} demo]
  {Importance of Many Particle Correlations to the Collective Debye-Waller Factor in a Single-Particle Activated Dynamic Theory of the Glass Transition}

\abbreviations{IR,NMR,UV}

\begin{document}



\begin{abstract}
  We theoretically study the importance of many body correlations on the collective Debye Waller (DW) factor in the context of the Nonlinear Langevin Equation (NLE) single particle activated dynamics theory of glass transition and its extension to include collective elasticity (ECNLE theory). 
This microscopic force-based approach envisions structural alpha relaxation as a coupled local-nonlocal process involving correlated local cage and longer range collective barriers.
The crucial question addressed here is the importance of the deGennes narrowing contribution versus a literal Vineyard approximation for the collective DW factor that enters the construction of the dynamic free energy in NLE theory. 
While the Vineyard-deGennes approach-based NLE theory and its ECNLE theory extension yields predictions that agree well with experimental and simulation results, use of a literal Vineyard approximation for the collective DW factor massively overpredicts the activated relaxation time. 
The current study suggests many particle correlations are crucial for a reliable description of activated dynamics theory of model hard sphere fluids.
\end{abstract}

\section{I. Introduction} 
Understanding glassy dynamics from a microscopic statistical mechanical perspective remains a multifaceted challenge of soft condensed matter physics, material science and chemical physics with enormous scientific and technological relevance~\cite{angell_2000_relaxation,berthier_2011_theoretical,ediger_2012_perspective,ronaldgarylarson_1999_the,wolfganggotze_2012_complex,dhont_2005_an}. 
The microscopic ideal Mode-Coupling Theory (MCT) idea of projecting real forces onto pair structural correlations to construct dynamical constraints for local caging and collective dynamic density fluctuations has been successful in understanding some aspects of glassy dynamics in model hard sphere fluids with growing density ~\cite{wolfganggotze_2012_complex,kirkpatrick_1987_connections,reichman_2005_modecoupling,janssen_2018_modecoupling}. 
However, ideal MCT suffers from an unphysical diverging relaxation time at modest packing fractions much below the experimental range of kinetic vitrification~\cite{janssen_2018_modecoupling,charbonneau_2014_hopping}. 
This suggests that while MCT can describe the initial stages of slowing down of the structural relaxation time, and perhaps short/intermediate time dynamics, the lack of consideration of activated dynamics plays a crucial role for the long-time structural or alpha relaxation process.

The Non-linear Langevin Equation (NLE) theory, first proposed nearly two decades ago~\cite{saltzman_2003_transport,schweizer_2003_entropic,schweizer_2005_derivation}, adopts MCT motivated ideas for quantifying caging constraints using equilibrium pair structural information, but goes beyond ideal MCT to critically include the effects of thermal noise driven fluctuations and activated hopping at a single particle level.
The more recent extension of NLE theory to include longer range collective elasticity effects  (Elastically Collective NLE, ECNLE)~\cite{mirigian_2013_unified,mirigian_2014_elastically,mirigian_2014_elastically2} is based on the idea of coupled local cage and nonlocal beyond the cage scale relaxation.
This theory has been successful in understanding the observable ~5-6 orders of magnitude relaxation in metastable hard sphere fluids.
However, since the classic formulations of MCT, NLE and ECNLE theories are based on projecting real forces onto equilibrium pair correlation function, possible dynamic consequences beyond that of pair structure are lost.
As a result, Lennard-Jones (LJ) and Weeks-Chandler-Anderson (WCA) potentials having very similar pair structure under certain thermodynamic conditions will be predicted to have very similar dynamics based on the unmodified ECNLE or MCT theories.
Isochoric model simulation studies have shown~\cite{berthier_2009_nonperturbative,berthier_2010_critical,berthier_2011_jcp,berthier_2011_testing,berthier_2011_theoretical} this to not be the case, at least at modestly high liquid-like densities, with the LJ fluid showing much slower relaxation due to attractive forces.
Importantly,  these effects get smaller and eventually nearly vanish at high enough fluid densities and perhaps under isobaric conditions where density grows with cooling. 

The difference in the dynamics of WCA and LJ fluid has been addressed in the ECNLE theory framework  based on retaining explicit information about the real Newtonian pair forces that goes beyond use of the projection approximation. 
Both real Newtonian forces and collective elastic effects have been shown to be crucial to explain the observed dynamical differences at the pair correlation functional level between WCA and LJ dense fluids~\cite{dell_2015_microscopic}. 
Recent machine learning (ML) studies have found that proper accounting for the small differences in LJ and WCA pair structural information can distinguish the observable difference in their dynamics~\cite{landes_2020_attractive}. 
Subsequent work in refs.~\cite{nandi_2017_role,saha_2019_a,nandi_2021_microscopic} argued a “modified” NLE theory that still retains the formal projection approximation can also differentiate the dynamical changes between LJ and WCA without any collective elastic contribution. 
Otherwise put, this ‘modified’ NLE theory can properly ‘weight’ structural information to predict substantial differences in dynamics from information about pair structure of LJ and WCA fluids.

The goal of the current article is to establish the reasons why the proposed simpler form of NLE theory in ref. ~\cite{nandi_2017_role,saha_2019_a,nandi_2021_microscopic} can apparently differentiate between LJ and WCA by comparing it with the original formulation of NLE theory in ref.~\cite{saltzman_2003_transport,schweizer_2003_entropic,schweizer_2005_derivation} for one-component HS fluids.
The results show there is a massive difference of relaxation timescales between the two versions of the theory which arises from the importance of the so-called deGennes narrowing contribution in the collective Debye-Waller (DW) factor or the dynamic propagator of collective force relaxation channel that is neglected in refs.~\cite{nandi_2017_role,saha_2019_a,nandi_2021_microscopic} but included in the original NLE theory of refs.~\cite{saltzman_2003_transport,schweizer_2003_entropic,schweizer_2005_derivation}. 
Hence the current article goes beyond just establishing the apparent distinction between the dynamics of LJ and WCA fluids to crucially identify the importance of properly modeling the collective dynamic propagator or DW factor in a single particle activated dynamics theory.

Section II briefly recalls elements of Naïve MCT (NMCT), NLE and ECNLE theories. 
The crucial elements of the proposed modified version of NLE theory in ref.~\cite{saha_2019_a} are also recalled, although aspects of it related to the use of a literal Vineyard approximation were already briefly discussed in the original NLE theory long ago by Saltzman and Schweizer~\cite{saltzman_2003_transport} and even earlier by Kirkpatrick and Wolynes~\cite{kirkpatrick_1987_connections}. 
Key quantities and the origin of dynamical differences of the alpha relaxation time predictions of the two versions of NLE theory are presented in Sec. III. Finally, we conclude in Sec. IV with a discussion and conclusions.

\section{II. Theoretical Background}
We briefly recall only the most relevant aspects of NMCT and NLE theories~\cite{saltzman_2003_transport,schweizer_2003_entropic,schweizer_2005_derivation} along with the ECNLE extension~\cite{mirigian_2013_unified,mirigian_2014_elastically,mirigian_2014_elastically2} for model hard sphere (HS) fluids of packing fraction$\phi=\frac{\pi}{6}\rho\sigma^3$, where $\rho$ is the particle number density and $\sigma$ is the particle diameter. 
\subsection{II. A. Ideal NMCT for HS fluids}
The starting point is a Generalized Langevin Equation (GLE) for ensemble-averaged tagged particle dynamics~\cite{saltzman_2003_transport,schweizer_2003_entropic,schweizer_2005_derivation}. 
The crucial quantity is the force-force time correlation function, which is computed based on naïve (single particle) Mode Coupling Theory (NMCT) as~\cite{kirkpatrick_1987_connections,saltzman_2003_transport,schweizer_2003_entropic,schweizer_2005_derivation},
\begin{equation}
     K(t)=\langle \vec{f}_0(t).\vec{f}_0(0)\rangle = \frac{1}{3\beta^{2}}\int \frac{d\vec{k}}{(2\pi)^3}\rho k^2C(k)^2S(k) \Gamma_s(k,t)\Gamma_c(k,t) 
\end{equation}
where $\beta=(k_BT)^{-1}$ is the inverse thermal energy, $\Gamma_s(k,t)=\langle e^{i\vec{k}.(\vec{r}(t)-\vec{r}(0))} \rangle$ and $\Gamma_c(k,t)=S(k,t)/S(k)$ are the (normalized to unity at $t=0$) single and collective dynamic structure factors or propagators, respectively. 
For a literal (ideal) glass, the dynamic propagators do not decay to zero as $t\to\infty$ and a localized state is described by a Debye-Waller factor in Eq. (1) that approximates single particle density fluctuations as Gaussian quantities with a characteristic localization length $r_L$. 
The ``Vineyard'' approximation~\cite{kirkpatrick_1987_connections,schweizer_2003_entropic,jeanpierrehansen_2006_theory} replaces collective density fluctuations by its single particle analog, resulting in,  
\begin{equation}
    \Gamma_c(k,t\to\infty)\approx \Gamma_s(k,t\to\infty)=\exp{\Big(-\frac{k^2 r_L^2}{6}\Big)}
\end{equation}
While this approximation is valid strictly in the high-wavevector ($k\to\infty$) limit, collective density fluctuations strongly differ compared to the single particle analog, especially at length-scales that define the local structural cage order. 
The origin is,  $S(k)>1$ for $k\sim k^\star$ (where $k^\star$ is the primary peak of static structure factor) and $S(k)< 1$ or even $\ll 1$ for “small” $k$.  
Quasi-elastic scattering experiments provide relaxation times of the microscopic structure with a characteristic length scale. 
DeGennes predicted~\cite{gotze_1992_relaxation,vanmegen_1998_measurement,kleban_1974_toward,degennes_1959_liquid} the $k$-dependence of the alpha relaxation time of collective or coherent dynamic structure factor, $S(k,t)$, exhibits a maximum around the first peak of the static structure factor, $S(k)$. 
The configuration of ‘aggregated collection’~\cite{saito2016slow} of particles is relatively more stable due to ‘cage formation’ compared to smaller or larger length scales. 
The characteristic dynamic slowdown near the range of wavevectors where $S(k)$ peaks ($k=k^\star$) is known as the deGennes narrowing effect and has been observed in neutron and x-ray scattering experiments on a wide variety of liquids, colloids, polymers, and proteins~\cite{sobolev_2016_de,wu_2018_atomic}.
This well known physics motivated~\cite{schweizer_2003_entropic} incorporating the deGennes narrowing effect in a modified Gaussian-type contribution for the collective Debye Waller factor in NMCT, which is no longer Gaussian but, has many particle corrections associated with S(k) or static collective density fluctuations. 
With this Vineyard-deGennes approximation the collective propagator is,
\begin{equation}
    \Gamma_c(k,t\to\infty)\approx \Gamma_s\Big(\frac{k}{\sqrt{S(k)}},t\to\infty\Big)=\exp{\Big(-\frac{k^2 r_L^2}{6S(k)}\Big)}
\end{equation}
The first approximate equality in a time-dependent context is called the “Skold approximation”~\cite{schweizer_2003_entropic} and considers static many body effects from $S(k)$ but assumes explicit dynamical processes can be modeled per single particle self-motions.
Hence, the deviations from the Vineyard approximation for the collective propagator depends on length scale, the importance of which are not a priori known with respect to questions of “glass physics”. 
While aspects of this physically intuitive choice of many-body corrections were briefly analyzed in ref.~\cite{schweizer_2003_entropic}, the goal of the present article is to present a detailed analysis of it compared to a literal Vineyard approach. 
Our results  clearly establish that the Vineyard-deGennes approximation is ‘much better’ for the collective propagator, consistent with the physical arguments that lead to its initial use in NCMT and NLE theories~\cite{schweizer_2003_entropic}.

The NMCT self-consistent localization equation follows from the condition $\frac{1}{2} \langle \vec{f}_0(t\to\infty).\vec{f}_0(0)\rangle r_L^2=\frac{3k_BT}{2}$, yielding ~\cite{kirkpatrick_1987_connections,schweizer_2003_entropic,schweizer_2005_derivation},
\begin{equation}
    \frac{1}{r_L^2} = \frac{\rho}{18\pi^2}\int_0^\infty dk k^4 C(k)^2 S(k)\exp{\Big(-\frac{k^2 r_L^2}{6}(1+S^{-1}(k))\Big)}
\end{equation}
Localization emerges (finite $r_L$) at $\phi_c=0.432$ for hard spheres using PY closure~\cite{schweizer_2003_entropic}.
Without the deGennes narrowing contribution (i.e., literal Vineyard approximation), the exponential factor inside the integrand of Eq. (4) becomes $\exp{\big(-k^2r_L^2/3\big)}$, and using the PY closure, we find, $\phi_c=0.334$, hence a localized state emerges at much lower packing fraction. 

\subsection{II.B. Activated Dynamic NLE Theory}
NLE theory for activated single particle dynamics is formulated in terms of stochastic trajectories controlled by ‘dynamic free energy’ that is constructed in terms of dynamic constraints motivated by NMCT ideas. 
The angularly averaged instantaneous displacement of a tagged single particle, $r(t)$, follows the NLE equation~\cite{schweizer_2003_entropic,schweizer_2005_derivation},
\begin{equation}
    \zeta_s\frac{dr(t)}{dt} = -\frac{\partial F_{\text{dyn}}(r(t))}{\partial r(t)}+\delta f(t)
\end{equation}
where the fluctuating thermal noise is related to the short time friction constant following fluctuation dissipation theorem as, $\langle \delta f(t).\delta f(0)\rangle =2k_BT\zeta_s\delta(t)$. 
Explicitly, the dynamic free energy is given as~\cite{schweizer_2003_entropic,nandi_2017_role},
\begin{eqnarray}
 \beta F_{\text{dyn}}^{\text{VdG}}(r)=-3\ln{(r)}-\int \frac{d\vec{k}}{(2\pi)^3}\frac{\rho           C(k)^2S(k)}{1+S^{-1}(k)}
        e^{-\frac{k^2r^2}{6}(1+S^{-1}(k))} \hspace{0.8cm} \label{eq:dg}\\
 \beta F_{\text{dyn}}^{\text{V}}(r)=-3\ln{(r)}-\int \frac{d\vec{k}}{(2\pi)^3}\frac{\rho C(k)^2S(k)}{2}
        e^{-\frac{k^2r^2}{3}}\hspace{2cm} \label{eq:ndg}
\end{eqnarray}
where the superscript `VdG' and `V' in Eq. (6) signifies Vineyard with the deGennes narrowing contribution or the literal Vineyard approximation in the dynamic free energy, respectively. 
The first term in Eq(6) favors particle delocalization and hence the fluid state, while the second term, $\beta F_{\text{ex}} (r)$ is an excess free energy associated with caging and favors particle localization. 
Note that Eq. (6b) is obtained by setting $(1+S^{-1} (k))=2$ in Eq. (6a), which is true only in the limit of high wavevectors since $\displaystyle \lim_{k\to \infty} S(k)\to 1$.
The integrand in the excess free energy of Eq (4) defines a length scale (or wavevector) dependent mean square force correlation or vertex that quantifies dynamic constraints on a length scale $\sim 2\pi/k$. 
Differences between the two vertex functions are analyzed later in the article.
\begin{figure}[ht!]
    \centering
    \includegraphics[scale=0.45]{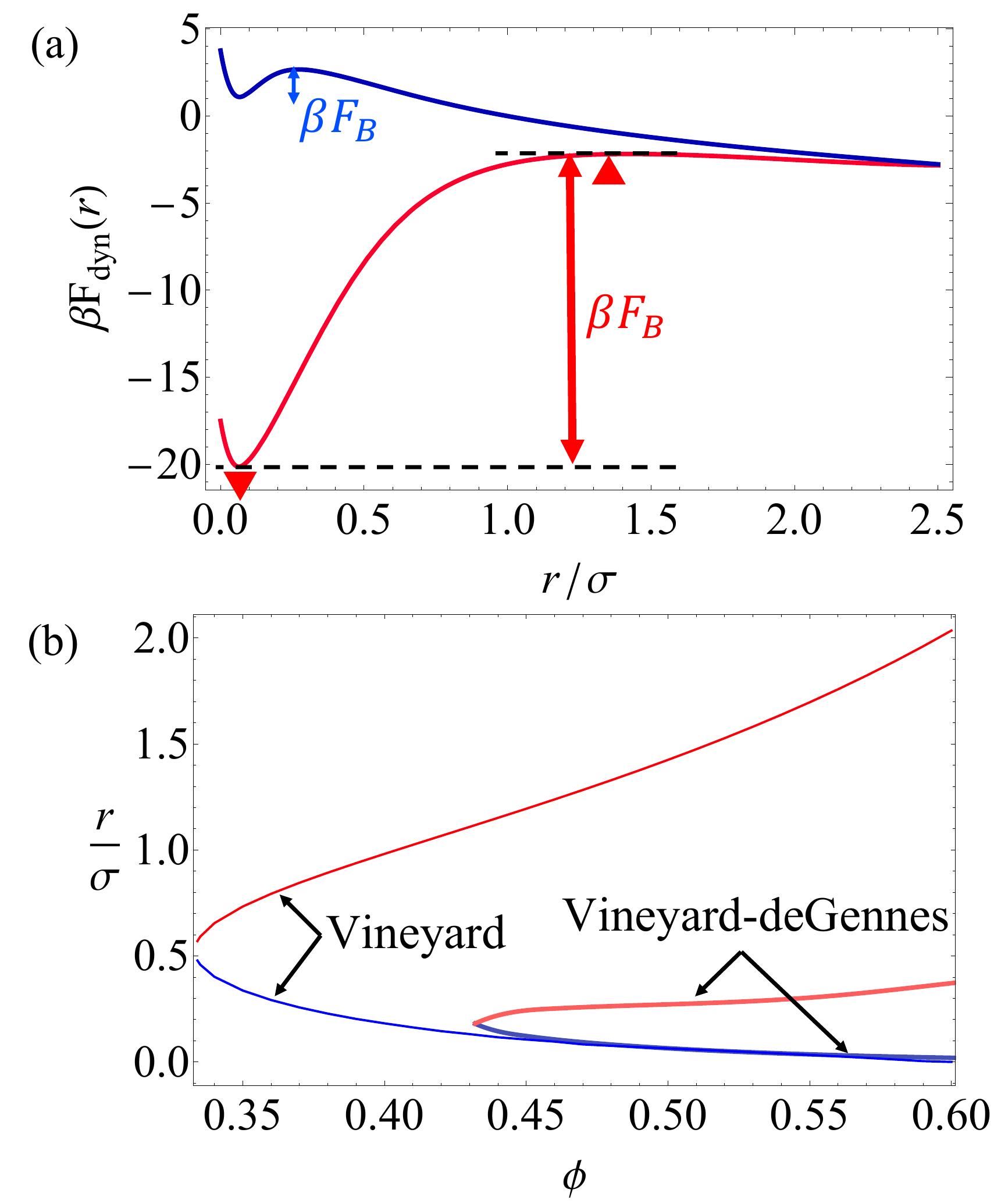}
    \caption{(a) Dynamic free energy for a rather low fixed packing fraction of $\phi=0.50$ is shown for the VdG (blue) and V (red) theories. The localization length and barrier location are shown for the V theory in upside down and regular triangles, respectively. Local cage barriers are also shown for both the theories in red ($\sim 18k_B T$) and blue ($\sim 2k_B T$) respectively. (b) Characteristic length scales: localization length (blue) and barrier location (red). The results of the VdG and V theories are represented as thick and thin curves respectively. }
    \label{fig:figure1}
\end{figure}

Beyond $\phi>\phi_c$, the dynamic free energy has a minimum at $r=r_L$ and a barrier at $r_B$ corresponding to a jump distance of $\Delta r_j = r_B-r_L$. 
A representative dynamic free energy and the characteristic length-scales for both the versions of NLE theory are shown in Fig. 1a. 
We note that the literal Vineyard based theory predicts a similar localization length as the VdG theory for $\phi\in (0.50,0.60)$ as shown in blue curves in Fig.1b. 
This can be understood since the relevant wavevectors ($k\sim k_r$) for localization obey $k_r r_L\sim 1$ or $k_r \sigma \gg 1$. 
Since the localization length is typically 1-2 orders of magnitude smaller than a particle diameter, $r_L \ll \sigma$, the Vineyard approximation is expected to be reasonable (dominated by high wavevectors) at the level of ideal NMCT.
This also means that the curvature of dynamic free energy at the localization length, $K_0\sim r_L^{-2}$, is similar for both versions of the theory. 
However, predictions are very different for barrier locations as shown in red in Fig. 1b, since the relevant wavevector for barrier location, $k_r r_B\sim 1$ implies $k_r \sigma \gg 1$ which is not satisfied as $r_B\sigma \ll 1$ does not hold and represents an internal inconsistency.  
This results in barrier locations often beyond a particle diameter ($r_B>\sigma $ beyond $\phi=0.45$) for Vineyard approximation theory (not true for the original NLE theory with VdG as shown in Fig. 1b), and hence a hopping event involving barrier escape involves a much larger amplitude ‘jump’. 
Physically, the latter suggests, as we show below, the Vineyard based version of NLE theory will predict relaxation times much longer than the analogous VdG theory.
\begin{figure}[ht!]
    \centering
    \includegraphics[scale=0.26]{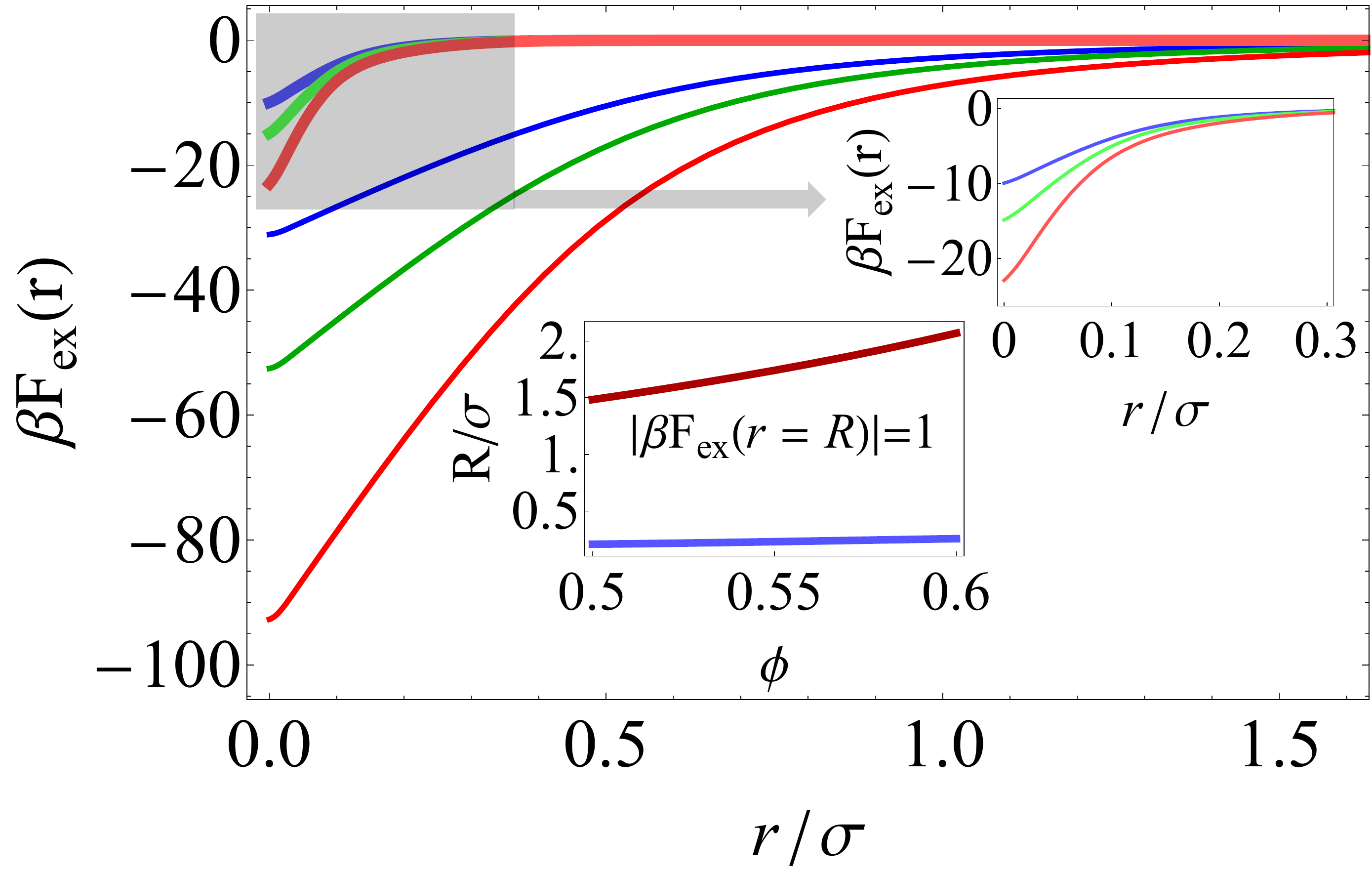}
    \caption{Excess dynamic free energy in thermal energy units as a function of displacement in units of the hard sphere diameter for three different packing fractions of 0.50 (blue), 0.55 (green) and 0.60 (red) for Vineyard (thin solid, zoomed in the top inset) and Vineyard plus deGennes (thick solid lines) approximation for the collective dynamic propagator. (Bottom inset) Lengthscale defined from $|\beta F_{ex}(r=R)|=1$ is plotted as a function of packing fraction for Vineyard (red) and Vineyard plus deGennes (blue) theories.}
    \label{fig:figure2}
\end{figure}

The excess part of the dynamic free energy that favors localization in Eq. (6) is explicitly shown in Fig. 2 for several packing fractions for both versions of the NLE theory.
To understand the lengthscale on which the excess free energy decays to zero, we follow ref.~\cite{schweizer_2003_entropic} and define a length from $|\beta F_{\text{ex}} (r=R)|=1$. 
The dimensionless lengthscale $R$ is plotted as a function of packing fraction in the inset of Fig.2. 
While the Vineyard-deGennes based NLE theory results show the range of the excess dynamic free energy is of order of $\sim 1/4^{\text{th}}-1/3^{\text{rd}}$ of a particle diameter, ignoring the deGennes correction leads to a massively longer range of  $\sim 1.5-2$ particle diameters which does not seem physically intuitive.
Another huge difference is the quantitative values of excess dynamic free energy at $r=0$. 
For $\phi=0.50$ and $0.60$ one has $\beta F_{\text{ex}}(r=0)= -10$ and $-23$, respectively, for the original NLE theory, while for the same packing fractions Vineyard NLE theory predicts values of $-32$ and $-93$, respectively.

\section{III. Theory Predictions}
\subsection{III. A. Alpha Relaxation Time}
The alpha relaxation time in NLE theory is computed as the mean first passage time from the Kramers theory of activated barrier hopping and given as~\cite{schweizer_2003_entropic,schweizer_2005_derivation},
\begin{equation}
    \frac{\tau_{\text{hop}}^{\text{NLE}}}{\tau_s}=\sigma^{-2}\int_{r_L}^{r_B}e^{\beta F_{\text{dyn}}(r_1)}dr_1\int_{r_L}^{r_1}e^{-\beta F_{\text{dyn}}(r_2)}dr_2 
    \approx \frac{2\pi}{\sqrt{K_0K_B}}e^{\beta F_B}
\label{eq:taunle}
\end{equation}
where $K_0$ and $K_B$ are the harmonic spring constants of the dynamic free energy at $r=r_L$ and $r=r_B$ respectively and the final approximate relation holds when the dynamic free energy barrier, $\beta F_B$ is beyond several thermal energy units.
Results for the local barrier as a function of packing fraction are shown in the Appendix.
The numerical results for the relaxation time employ the full integral in Eq.\ref{eq:taunle} and time is expressed in units of the short timescale related to Fickian diffusion of the system as $\tau_s=\beta \zeta_s\sigma^2$.

\begin{figure}
    \centering
    \includegraphics[scale=0.25]{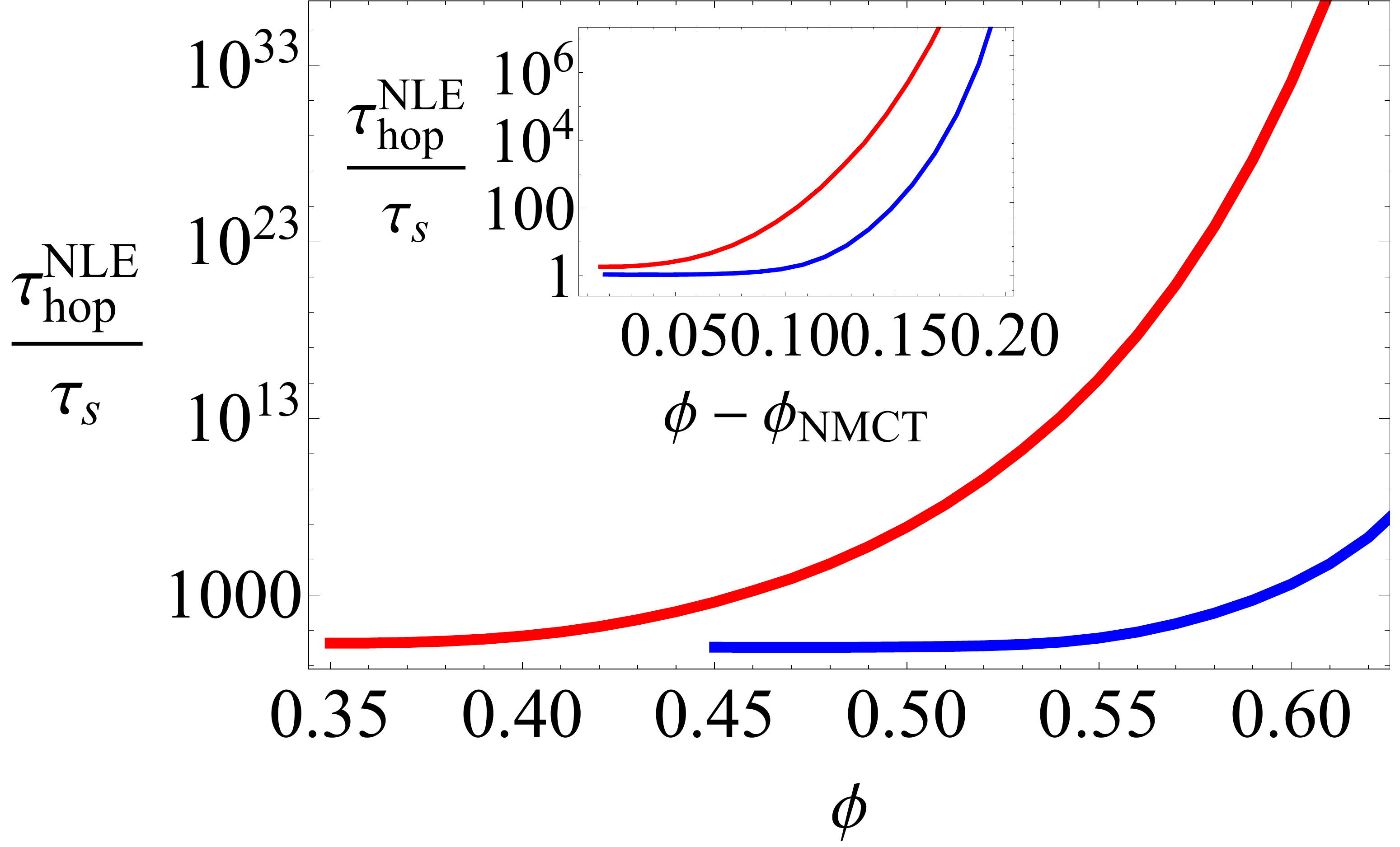}
    \caption{Mean first passage time from Kramers theory are shown for the two versions of the theory in red (no deGennes) and blue (deGennes). A zoomed in version for a smaller range of relaxation time is shown in the inset as a function of distance from ideal NMCT transition packing fraction, $(\phi-\phi_{NMCT})$. The two versions of NLE theory do not collapse. }
    \label{fig:figure3}
\end{figure}

Figure 3 plots the NLE theory relaxation times as a function of packing fractions for both versions of theory. While in the range $\phi\in (0.50-0.60)$ the NLE theory with the deGennes contribution predicts a modest $\sim 3.5$ orders of slowdown, without the deGennes contribution more than $\sim 20$ orders of magnitude slowdown in relaxation dynamics is predicted.
Hence, while the two approaches are similar in spirit, their activated relaxation time predictions are very different (cf. local barrier in Appendix). 
Ignoring the deGennes contribution massively overpredicts dynamical slowing down. Empirically horizontally shifted (a commonly adopted procedure when comparing MCT to experiment or simulation) NLE results are shown in the inset as a function of distance from the ideal NMCT transition of $0.432$ (blue, Vineyard-deGennes) or $0.334$ (red, Vineyard). 
Vineyard NLE theory also predicts much stronger growth of the relaxation time in this representation.  
The two set of curves from the different NLE theories do not collapse based on any naive horizontal shift.

\subsection{III. B. Collective Elastic Contribution}
It has been recently argued that the alpha relaxation is a coupled local-nonlocal activated process involving both a local cage barrier (as shown in Fig. 1.a) and a collective elastic barrier ($\beta F_{\text{el}}$) leading to the Elastically Collective NLE (ECNLE) theory~\cite{mirigian_2013_unified,mirigian_2014_elastically,mirigian_2014_elastically2}. 
The central idea relies on the fact that a local cage scale hopping requires tiny displacements of surrounding particles leading to an effective cage expansion that introduces an additional barrier for the activated event to occur~\cite{mirigian_2014_elastically}. 
The elastic barrier is calculated as,
\begin{equation}
    \beta F_{el} = 12\phi K_0\Delta r_{eff}^2 \left(\frac{r_{cage}}{\sigma}\right)^3
\end{equation}
where, $K_0=3k_BT/r_L^2$ is the harmonic spring constant or curvature of the dynamic free energy at its local minimum;
$\Delta r_{\text{eff}}=\frac{3}{32}\frac{\Delta r_j^2}{r_{\text{cage}}}$ is the effective amplitude (typically small, of order $r_L$ or smaller) of the elastic displacement field at the cage surface. 
Hence, the alpha relaxation process is a coupled local-nonlocal activated event involving both an elastic and local barrier, which modifies Eq.7 as, $\tau_\alpha = \tau_{\text{NLE}}\exp{(\beta F_{\text{el}})}$. 
From ref.~\cite{mirigian_2014_elastically} and Eq (9), the elastic barrier is predicted to grow strongly as the fourth power of jump distance ($\beta F_{\text{el}}\sim K_0 \Delta r_j^4$), where $\Delta r_{j}=r_B-r_L$. 
Considering a modest packing fraction of $\sim 0.50$, typical values of jump distance for NLE theory with the deGennes contribution is sensibly small, $\sim 0.3\sigma$, yielding an elastic barrier of $\sim 0.20k_BT$. 
In contrast, the jump distance for the literal Vineyard formulation is a factor of more than $4$ larger $\sim 1.4\sigma$. 
Since the harmonic spring constant $K_0=3k_BT/r_L^{2}$ predicted by the two formulations of the collective DW are close to each other to zeroth order, this gives an elastic barrier that is $\sim 4^4\sim 250$ times larger for ECNLE theory without the deGennes narrowing contribution or, close to $\sim 50k_BT$, leading to an enormous slowdown of an additional $\sim 20$ orders of magnitude in alpha time. 
Results for the packing fraction dependence of elastic barrier is presented in the Appendix.

The predictions of ECNLE theory with the deGennes contribution~\cite{mirigian_2014_elastically,mirigian_2014_elastically2} have been shown to agree well with experimental and simulation data for polydisperse hard-sphere systems~\cite{mirigian_2014_elastically2} and monodisperse hard spheres~\cite{mei2020thermodynamics,mei2021experimental}.
The theory has been extended to study molecular and polymeric liquids~\cite{mirigian_2015_dynamical} and other more complex colloidal systems (such as soft~\cite{ghosh_2019_linear}, non-spherical, gel forming~\cite{ghosh_2019_microscopic,ghosh_2020_the} fluids or suspensions).
A  zeroth order accounting of dynamic heterogeneity has also been carried out, from both a static disorder point of view~\cite{xie_2020_a} or a stochastic trajectory perspective~\cite{saltzman2006activated}.
The role of attractive forces, a re-entrant glass transition~\cite{ghosh_2020_microscopic}, and the effect of deformation (stress, strain) have also been addressed~\cite{ghosh_2020_microscopic}.
Without the deGennes contribution in the collective propagator, baseline predictions for model hard sphere systems are much worse and seem unphysical in a practical sense, and hence adopting a pure Vineyard approximation cannot be a reliable description of more complicated systems.

To note, calculations of the original elastic barrier are constructed using continuum mechanics analysis and Einstein model of a solid~\cite{mirigian_2014_elastically}, that considers harmonic displacements for small jump distances.
While the same is no longer true for the pure Vineyard approximation, surrounding particles still need to displace to accommodate the much larger hopping process and continuum elastic picture still applies in a dense system.
Hence, even though the larger magnitude of jump distances pose a difficulty on the validity of Einstein solid approximation, to a zeroth order it provides a good qualitative estimate of an elastic barrier.

\subsection{III. C. Origin of Dynamical Differences}
\begin{figure}[ht!]
    \centering
    \includegraphics[scale=0.16]{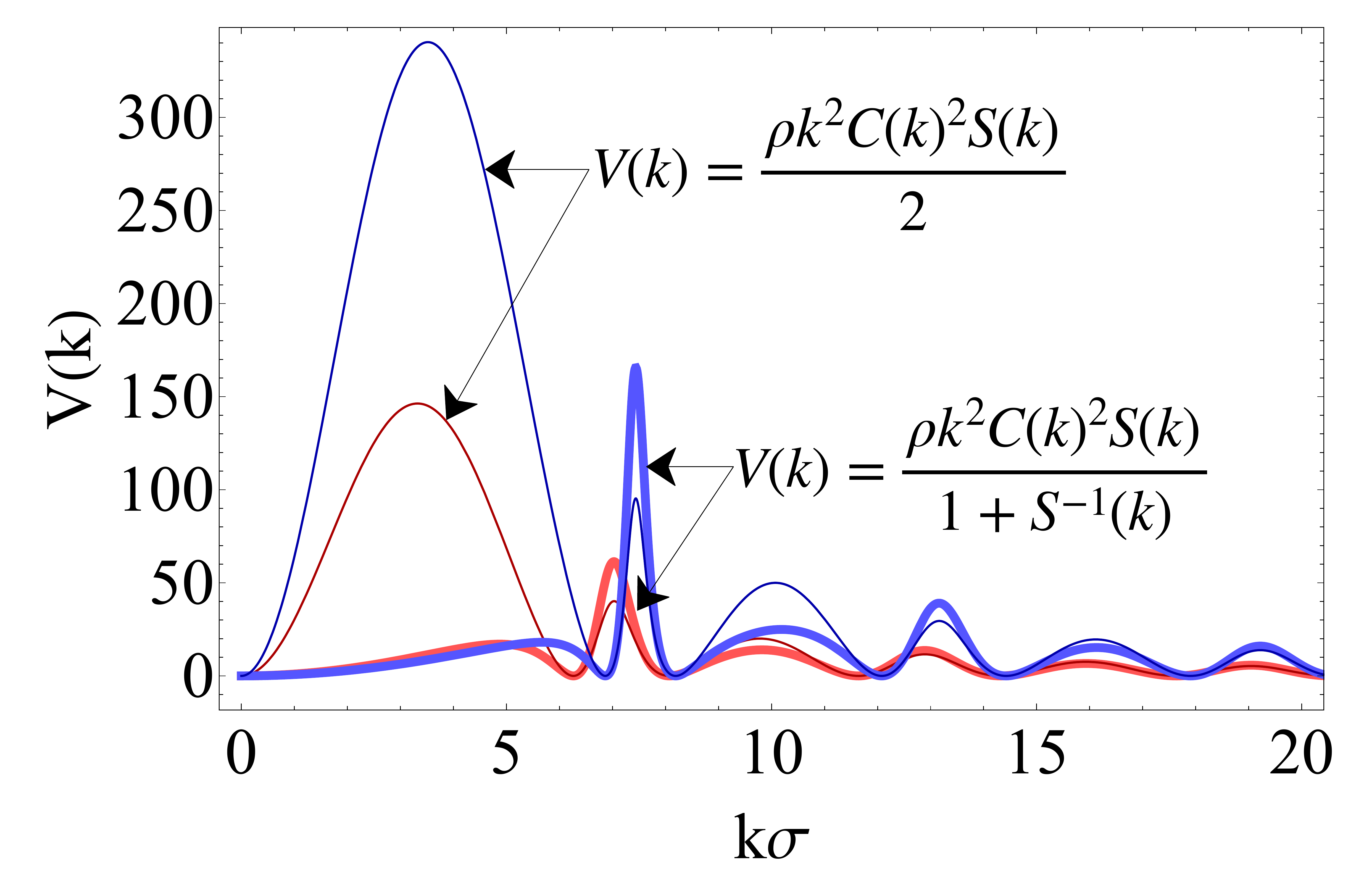}
    \caption{Vertex defined by the two different approaches are plotted for two packing fractions of 0.50 (red) and 0.60 (blue). With deGennes narrowing contribution is shown in thick red and blue lines while ndG theory is shown in thin red and blue plots.}
    \label{fig:figure4}
\end{figure}
To understand the origin of the dynamical differences between the two versions of the theory (at the NLE or ECNLE levels, since both make predictions for the required dynamical properties from knowledge of the same dynamic free energy), we plot the integrand in Eq.6 without the exponential contribution in the caging term, i.e., we define the wavevector dependent vertex,
\begin{eqnarray}
 V^{\text{VdG}}(k)=\frac{k^2C(k)^2S(k)}{1+S^{-1}(k)}\hspace{1 cm}\\
 V^{\text{V}}(k)=\frac{k^2C(k)^2S(k)}{2}\hspace{1 cm}
\end{eqnarray}
The terms in Eq.8 are plotted as a function of dimensionless wave-vector for two packing fractions of 0.50 (red) and 0.60 (blue). 
While $V^{VdG} (k)$ has a maximum that corresponds well with the location of the cage-coherence peak of the static structure factor (i.e., primary peak of $S(k)$ at $k=k^\star$) as physically expected, the maximum in $V^V (k)$ occurs at a wave-vector $k<k^\star$, indicating the primary contribution to the integrand of $\beta F_{dyn}^V (r)$ comes from larger than the cage lengthscale. 
The latter behavior seems physically and mathematically inconsistent with the use of the Vineyard high-k dominance approximation for the collective DW factor. 
The denominator in Eq.(8a) is crucially responsible for suppressing large contributions at small wave-vectors since for $k \ll k^\star$ one has $S(k) \ll 1 $, which is completely ignored in the Vineyard approximation leading to massive overprediction of barriers and relaxation times.

It is also important to note that the literal Vineyard NLE theory approach still constructs a dynamic free energy from solely the two-point structural correlations, but eventually massively overpredicts the dynamics of model HS systems, hence it is much more sensitive to small changes in structure.
Hence, the predictions of a difference in dynamics of LJ and WCA fluids that differ structurally (two-point correlations) by small amounts originate from the literal Vineyard approximation in ref.~\cite{nandi_2017_role,saha_2019_a,nandi_2021_microscopic}.
However, this finding emerges from a theory that massively overpredicts activated relaxation times by many orders of magnitude beyond what any simulations or experiments find~\cite{mirigian_2014_elastically2} and exhibit a different dependence on fluid packing fraction than the original NLE theory.

\section{IV. Summary and Discussion}
We have analyzed in depth two different implementations of a microscopic force based activated dynamic theory of model hard sphere glassy liquids. 
While both versions build on the ideal MCT ideas to construct the dynamical constraints based static two-point correlation function structural input, the collective dynamic propagator in the force-force time correlation either ignores or includes contribution of static structure in the so-called Vineyard or Vineyard-deGennes approximation, respectively. 
Our analysis shows use of the literal Vineyard approximation leads to a massive overprediction of the alpha time of a HS liquid, while the Vineyard-deGennes approximation-based theory predictions have been shown to work very well per literature studies~\cite{mirigian_2014_elastically2,mirigian_2015_dynamical,ghosh_2019_linear,xie_2020_a,mei2020thermodynamics,mei2021experimental,ghosh_2019_microscopic,ghosh_2020_microscopic,ghosh_2020_the}. 
Hence, any activated dynamics theory such as analyzed here that employs the Vineyard approximation is almost guaranteed to overestimate dynamics and (unphysically) amplify small structural differences such as exist for the WCA and LJ fluids.  
We then conclude that claims~\cite{nandi_2017_role,saha_2019_a,nandi_2021_microscopic} that using a simpler Vineyard NLE theory can explain the differences between WCA and LJ fluids observed in simulation based on small differences in pair correlation function input is not reliable.

\appendix 
\section*{Appendix: Local and Elastic Barrier}

The local cage and collective elastic barrier from both versions of the presented NLE and ECNLE theories are plotted as a function of packing fraction in log-linear fashion in Fig. 5. 
Perhaps surprisingly, although the absolute magnitudes of local and elastic barriers from the two versions of presented theories differ, they both follow a roughly exponential growth with packing fraction in the ‘high’-$\phi$ region as shown by the ‘parallel’ dashed black lines in Fig. 5.
This behavior is no doubt a consequence of using the same structural pair correlation function input to quantify the kinetic constraints or dynamic vertex in the dynamical free energy. 
No simple horizontal and/or vertical shift can overlay the different cage and elastic barriers as predicted by the V or VdG approximations.
\begin{figure}[ht!]
    \centering
    \includegraphics[scale=0.85]{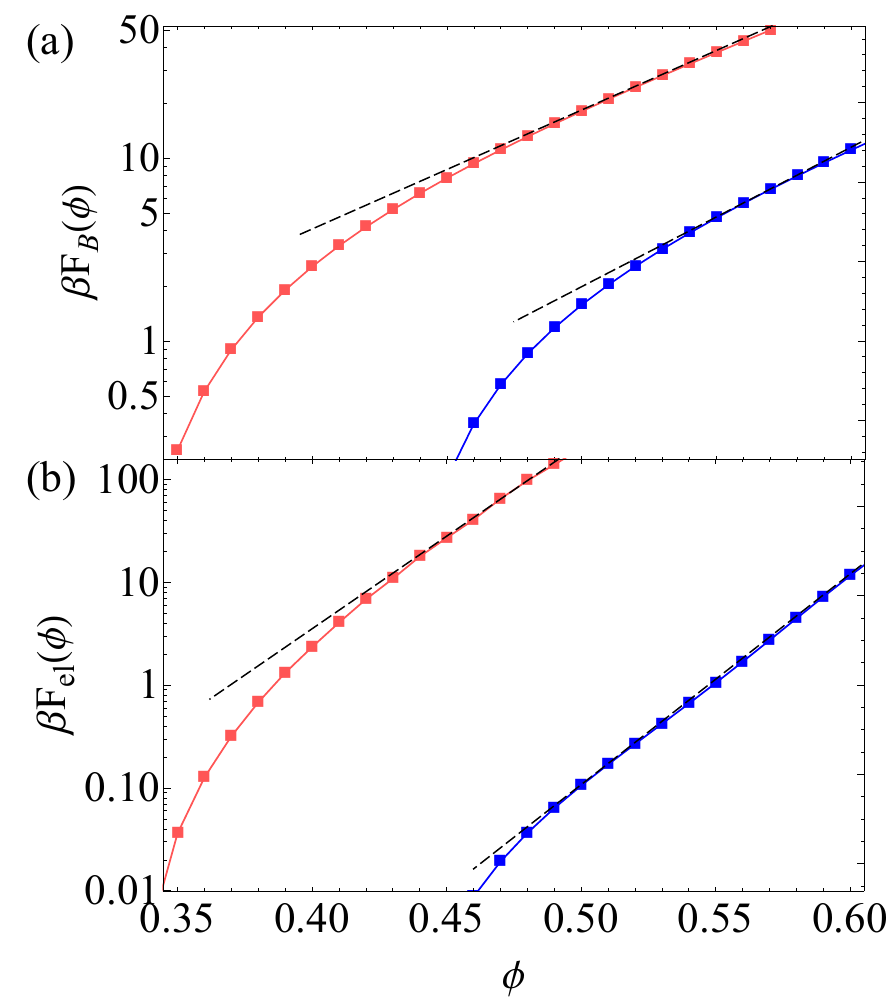}
    \caption{Local cage (panel (a)) and collective elastic (panel (b)) barrier as a function of packing fraction with (blue) and without (red) the deGennes correction. The black dashed lines demonstrate $\beta F\sim \exp{(a\phi)}$ dependence of barriers on packing fraction (a is a constant) in the ‘high’ packing fraction limit. }
    \label{fig:app1}
\end{figure}

\begin{acknowledgement}

I would like to thank Professor Kenneth S. Schweizer for helpful discussions and comments on the manuscript.\vspace{1 cm}

\end{acknowledgement}


\bibliography{achemso-demo}

\providecommand{\noopsort}[1]{}\providecommand{\singleletter}[1]{#1}
\providecommand{\latin}[1]{#1}
\makeatletter
\providecommand{\doi}
  {\begingroup\let\do\@makeother\dospecials
  \catcode`\{=1 \catcode`\}=2 \doi@aux}
\providecommand{\doi@aux}[1]{\endgroup\texttt{#1}}
\makeatother
\providecommand*\mcitethebibliography{\thebibliography}
\csname @ifundefined\endcsname{endmcitethebibliography}
  {\let\endmcitethebibliography\endthebibliography}{}
\begin{mcitethebibliography}{43}
\providecommand*\natexlab[1]{#1}
\providecommand*\mciteSetBstSublistMode[1]{}
\providecommand*\mciteSetBstMaxWidthForm[2]{}
\providecommand*\mciteBstWouldAddEndPuncttrue
  {\def\EndOfBibitem{\unskip.}}
\providecommand*\mciteBstWouldAddEndPunctfalse
  {\let\EndOfBibitem\relax}
\providecommand*\mciteSetBstMidEndSepPunct[3]{}
\providecommand*\mciteSetBstSublistLabelBeginEnd[3]{}
\providecommand*\EndOfBibitem{}
\mciteSetBstSublistMode{f}
\mciteSetBstMaxWidthForm{subitem}{(\alph{mcitesubitemcount})}
\mciteSetBstSublistLabelBeginEnd
  {\mcitemaxwidthsubitemform\space}
  {\relax}
  {\relax}

\bibitem[Angell \latin{et~al.}(2000)Angell, Ngai, McKenna, McMillan, and
  Martin]{angell_2000_relaxation}
Angell,~C.~A.; Ngai,~K.~L.; McKenna,~G.~B.; McMillan,~P.~F.; Martin,~S.~W.
  Relaxation in glassforming liquids and amorphous solids. \emph{Journal of
  Applied Physics} \textbf{2000}, \emph{88}, 3113--3157\relax
\mciteBstWouldAddEndPuncttrue
\mciteSetBstMidEndSepPunct{\mcitedefaultmidpunct}
{\mcitedefaultendpunct}{\mcitedefaultseppunct}\relax
\EndOfBibitem
\bibitem[Berthier and Biroli(2011)Berthier, and
  Biroli]{berthier_2011_theoretical}
Berthier,~L.; Biroli,~G. Theoretical perspective on the glass transition and
  amorphous materials. \emph{Reviews of Modern Physics} \textbf{2011},
  \emph{83}, 587--645\relax
\mciteBstWouldAddEndPuncttrue
\mciteSetBstMidEndSepPunct{\mcitedefaultmidpunct}
{\mcitedefaultendpunct}{\mcitedefaultseppunct}\relax
\EndOfBibitem
\bibitem[Ediger and Harrowell(2012)Ediger, and
  Harrowell]{ediger_2012_perspective}
Ediger,~M.~D.; Harrowell,~P. Perspective: Supercooled liquids and glasses.
  \emph{The Journal of Chemical Physics} \textbf{2012}, \emph{137},
  080901\relax
\mciteBstWouldAddEndPuncttrue
\mciteSetBstMidEndSepPunct{\mcitedefaultmidpunct}
{\mcitedefaultendpunct}{\mcitedefaultseppunct}\relax
\EndOfBibitem
\bibitem[Larson(1999)]{ronaldgarylarson_1999_the}
Larson,~R.~G. \emph{The structure and rheology of complex fluids}; Oxford
  University Press, 1999\relax
\mciteBstWouldAddEndPuncttrue
\mciteSetBstMidEndSepPunct{\mcitedefaultmidpunct}
{\mcitedefaultendpunct}{\mcitedefaultseppunct}\relax
\EndOfBibitem
\bibitem[Gotze(2012)]{wolfganggotze_2012_complex}
Gotze,~W. \emph{Complex dynamics of glass-forming liquids : a mode-coupling
  theory}; Oxford University Press, 2012\relax
\mciteBstWouldAddEndPuncttrue
\mciteSetBstMidEndSepPunct{\mcitedefaultmidpunct}
{\mcitedefaultendpunct}{\mcitedefaultseppunct}\relax
\EndOfBibitem
\bibitem[Dhont(2005)]{dhont_2005_an}
Dhont,~J. K.~G. \emph{An introduction to dynamics of colloids}; Elsevier,
  2005\relax
\mciteBstWouldAddEndPuncttrue
\mciteSetBstMidEndSepPunct{\mcitedefaultmidpunct}
{\mcitedefaultendpunct}{\mcitedefaultseppunct}\relax
\EndOfBibitem
\bibitem[Kirkpatrick and Wolynes(1987)Kirkpatrick, and
  Wolynes]{kirkpatrick_1987_connections}
Kirkpatrick,~T.~R.; Wolynes,~P.~G. Connections between some kinetic and
  equilibrium theories of the glass transition. \emph{Physical Review A}
  \textbf{1987}, \emph{35}, 3072--3080\relax
\mciteBstWouldAddEndPuncttrue
\mciteSetBstMidEndSepPunct{\mcitedefaultmidpunct}
{\mcitedefaultendpunct}{\mcitedefaultseppunct}\relax
\EndOfBibitem
\bibitem[Reichman and Charbonneau(2005)Reichman, and
  Charbonneau]{reichman_2005_modecoupling}
Reichman,~D.~R.; Charbonneau,~P. Mode-coupling theory. \emph{Journal of
  Statistical Mechanics: Theory and Experiment} \textbf{2005}, \emph{P05013},
  P05013\relax
\mciteBstWouldAddEndPuncttrue
\mciteSetBstMidEndSepPunct{\mcitedefaultmidpunct}
{\mcitedefaultendpunct}{\mcitedefaultseppunct}\relax
\EndOfBibitem
\bibitem[Janssen(2018)]{janssen_2018_modecoupling}
Janssen,~L. M.~C. Mode-Coupling Theory of the Glass Transition: A Primer.
  \emph{Frontiers in Physics} \textbf{2018}, \emph{6}, 97\relax
\mciteBstWouldAddEndPuncttrue
\mciteSetBstMidEndSepPunct{\mcitedefaultmidpunct}
{\mcitedefaultendpunct}{\mcitedefaultseppunct}\relax
\EndOfBibitem
\bibitem[Charbonneau \latin{et~al.}(2014)Charbonneau, Jin, Parisi, and
  Zamponi]{charbonneau_2014_hopping}
Charbonneau,~P.; Jin,~Y.; Parisi,~G.; Zamponi,~F. Hopping and the
  Stokes–Einstein relation breakdown in simple glass formers.
  \emph{Proceedings of the National Academy of Sciences} \textbf{2014},
  \emph{111}, 15025--15030\relax
\mciteBstWouldAddEndPuncttrue
\mciteSetBstMidEndSepPunct{\mcitedefaultmidpunct}
{\mcitedefaultendpunct}{\mcitedefaultseppunct}\relax
\EndOfBibitem
\bibitem[Saltzman and Schweizer(2003)Saltzman, and
  Schweizer]{saltzman_2003_transport}
Saltzman,~E.~J.; Schweizer,~K.~S. Transport coefficients in glassy colloidal
  fluids. \emph{The Journal of Chemical Physics} \textbf{2003}, \emph{119},
  1197--1203\relax
\mciteBstWouldAddEndPuncttrue
\mciteSetBstMidEndSepPunct{\mcitedefaultmidpunct}
{\mcitedefaultendpunct}{\mcitedefaultseppunct}\relax
\EndOfBibitem
\bibitem[Schweizer and Saltzman(2003)Schweizer, and
  Saltzman]{schweizer_2003_entropic}
Schweizer,~K.~S.; Saltzman,~E.~J. Entropic barriers, activated hopping, and the
  glass transition in colloidal suspensions. \emph{The Journal of Chemical
  Physics} \textbf{2003}, \emph{119}, 1181--1196\relax
\mciteBstWouldAddEndPuncttrue
\mciteSetBstMidEndSepPunct{\mcitedefaultmidpunct}
{\mcitedefaultendpunct}{\mcitedefaultseppunct}\relax
\EndOfBibitem
\bibitem[Schweizer(2005)]{schweizer_2005_derivation}
Schweizer,~K.~S. Derivation of a microscopic theory of barriers and activated
  hopping transport in glassy liquids and suspensions. \emph{The Journal of
  Chemical Physics} \textbf{2005}, \emph{123}, 244501\relax
\mciteBstWouldAddEndPuncttrue
\mciteSetBstMidEndSepPunct{\mcitedefaultmidpunct}
{\mcitedefaultendpunct}{\mcitedefaultseppunct}\relax
\EndOfBibitem
\bibitem[Mirigian and Schweizer(2013)Mirigian, and
  Schweizer]{mirigian_2013_unified}
Mirigian,~S.; Schweizer,~K.~S. Unified Theory of Activated Relaxation in
  Liquids over 14 Decades in Time. \emph{The Journal of Physical Chemistry
  Letters} \textbf{2013}, \emph{4}, 3648--3653\relax
\mciteBstWouldAddEndPuncttrue
\mciteSetBstMidEndSepPunct{\mcitedefaultmidpunct}
{\mcitedefaultendpunct}{\mcitedefaultseppunct}\relax
\EndOfBibitem
\bibitem[Mirigian and Schweizer(2014)Mirigian, and
  Schweizer]{mirigian_2014_elastically}
Mirigian,~S.; Schweizer,~K.~S. Elastically cooperative activated barrier
  hopping theory of relaxation in viscous fluids. I. General formulation and
  application to hard sphere fluids. \emph{The Journal of Chemical Physics}
  \textbf{2014}, \emph{140}, 194506\relax
\mciteBstWouldAddEndPuncttrue
\mciteSetBstMidEndSepPunct{\mcitedefaultmidpunct}
{\mcitedefaultendpunct}{\mcitedefaultseppunct}\relax
\EndOfBibitem
\bibitem[Mirigian and Schweizer(2014)Mirigian, and
  Schweizer]{mirigian_2014_elastically2}
Mirigian,~S.; Schweizer,~K.~S. Elastically cooperative activated barrier
  hopping theory of relaxation in viscous fluids. II. Thermal liquids.
  \emph{The Journal of Chemical Physics} \textbf{2014}, \emph{140},
  194507\relax
\mciteBstWouldAddEndPuncttrue
\mciteSetBstMidEndSepPunct{\mcitedefaultmidpunct}
{\mcitedefaultendpunct}{\mcitedefaultseppunct}\relax
\EndOfBibitem
\bibitem[Berthier and Tarjus(2009)Berthier, and
  Tarjus]{berthier_2009_nonperturbative}
Berthier,~L.; Tarjus,~G. Nonperturbative Effect of Attractive Forces in Viscous
  Liquids. \emph{Physical Review Letters} \textbf{2009}, \emph{103},
  170601\relax
\mciteBstWouldAddEndPuncttrue
\mciteSetBstMidEndSepPunct{\mcitedefaultmidpunct}
{\mcitedefaultendpunct}{\mcitedefaultseppunct}\relax
\EndOfBibitem
\bibitem[Berthier and Tarjus(2010)Berthier, and Tarjus]{berthier_2010_critical}
Berthier,~L.; Tarjus,~G. Critical test of the mode-coupling theory of the glass
  transition. \emph{Physical Review E} \textbf{2010}, \emph{82}, 031502\relax
\mciteBstWouldAddEndPuncttrue
\mciteSetBstMidEndSepPunct{\mcitedefaultmidpunct}
{\mcitedefaultendpunct}{\mcitedefaultseppunct}\relax
\EndOfBibitem
\bibitem[Berthier and Tarjus(2011)Berthier, and Tarjus]{berthier_2011_jcp}
Berthier,~L.; Tarjus,~G. The role of attractive forces in viscous liquids.
  \emph{The Journal of Chemical Physics} \textbf{2011}, \emph{134},
  214503\relax
\mciteBstWouldAddEndPuncttrue
\mciteSetBstMidEndSepPunct{\mcitedefaultmidpunct}
{\mcitedefaultendpunct}{\mcitedefaultseppunct}\relax
\EndOfBibitem
\bibitem[Berthier and Tarjus(2011)Berthier, and Tarjus]{berthier_2011_testing}
Berthier,~L.; Tarjus,~G. Testing ''microscopic'' theories of glass-forming
  liquids. \emph{The European Physical Journal E} \textbf{2011}, \emph{34},
  96\relax
\mciteBstWouldAddEndPuncttrue
\mciteSetBstMidEndSepPunct{\mcitedefaultmidpunct}
{\mcitedefaultendpunct}{\mcitedefaultseppunct}\relax
\EndOfBibitem
\bibitem[Dell and Schweizer(2015)Dell, and Schweizer]{dell_2015_microscopic}
Dell,~Z.~E.; Schweizer,~K.~S. Microscopic Theory for the Role of Attractive
  Forces in the Dynamics of Supercooled Liquids. \emph{Physical Review Letters}
  \textbf{2015}, \emph{115}, 205702\relax
\mciteBstWouldAddEndPuncttrue
\mciteSetBstMidEndSepPunct{\mcitedefaultmidpunct}
{\mcitedefaultendpunct}{\mcitedefaultseppunct}\relax
\EndOfBibitem
\bibitem[Landes \latin{et~al.}(2020)Landes, Biroli, Dauchot, Liu, and
  Reichman]{landes_2020_attractive}
Landes,~F.~P.; Biroli,~G.; Dauchot,~O.; Liu,~A.~J.; Reichman,~D.~R. Attractive
  versus truncated repulsive supercooled liquids: The dynamics is encoded in
  the pair correlation function. \emph{Physical Review E} \textbf{2020},
  \emph{101}, 010602(R)\relax
\mciteBstWouldAddEndPuncttrue
\mciteSetBstMidEndSepPunct{\mcitedefaultmidpunct}
{\mcitedefaultendpunct}{\mcitedefaultseppunct}\relax
\EndOfBibitem
\bibitem[Nandi \latin{et~al.}(2017)Nandi, Banerjee, Dasgupta, and
  Bhattacharyya]{nandi_2017_role}
Nandi,~M.~K.; Banerjee,~A.; Dasgupta,~C.; Bhattacharyya,~S.~M. Role of the Pair
  Correlation Function in the Dynamical Transition Predicted by Mode Coupling
  Theory. \emph{Physical Review Letters} \textbf{2017}, \emph{119},
  265502\relax
\mciteBstWouldAddEndPuncttrue
\mciteSetBstMidEndSepPunct{\mcitedefaultmidpunct}
{\mcitedefaultendpunct}{\mcitedefaultseppunct}\relax
\EndOfBibitem
\bibitem[Saha \latin{et~al.}(2019)Saha, Nandi, Dasgupta, and
  Bhattacharyya]{saha_2019_a}
Saha,~I.; Nandi,~M.~K.; Dasgupta,~C.; Bhattacharyya,~S.~M. A comparative study
  of a class of mean field theories of the glass transition. \emph{Journal of
  Statistical Mechanics: Theory and Experiment} \textbf{2019}, \emph{2019},
  084008\relax
\mciteBstWouldAddEndPuncttrue
\mciteSetBstMidEndSepPunct{\mcitedefaultmidpunct}
{\mcitedefaultendpunct}{\mcitedefaultseppunct}\relax
\EndOfBibitem
\bibitem[Nandi and Bhattacharyya(2021)Nandi, and
  Bhattacharyya]{nandi_2021_microscopic}
Nandi,~M.~K.; Bhattacharyya,~S.~M. Microscopic Theory of Softness in
  Supercooled Liquids. \emph{Physical Review Letters} \textbf{2021},
  \emph{126}, 208001\relax
\mciteBstWouldAddEndPuncttrue
\mciteSetBstMidEndSepPunct{\mcitedefaultmidpunct}
{\mcitedefaultendpunct}{\mcitedefaultseppunct}\relax
\EndOfBibitem
\bibitem[Hansen and Mcdonald(2006)Hansen, and
  Mcdonald]{jeanpierrehansen_2006_theory}
Hansen,~J.-P.; Mcdonald,~I.~R. \emph{Theory of simple liquids}; Elsevier /
  Academic Press, 2006\relax
\mciteBstWouldAddEndPuncttrue
\mciteSetBstMidEndSepPunct{\mcitedefaultmidpunct}
{\mcitedefaultendpunct}{\mcitedefaultseppunct}\relax
\EndOfBibitem
\bibitem[Gotze and Sjogren(1992)Gotze, and Sjogren]{gotze_1992_relaxation}
Gotze,~W.; Sjogren,~L. Relaxation processes in supercooled liquids.
  \emph{Reports on Progress in Physics} \textbf{1992}, \emph{55},
  241--376\relax
\mciteBstWouldAddEndPuncttrue
\mciteSetBstMidEndSepPunct{\mcitedefaultmidpunct}
{\mcitedefaultendpunct}{\mcitedefaultseppunct}\relax
\EndOfBibitem
\bibitem[van Megen \latin{et~al.}(1998)van Megen, Mortensen, Williams, and
  Müller]{vanmegen_1998_measurement}
van Megen,~W.; Mortensen,~T.~C.; Williams,~S.~R.; Müller,~J. Measurement of
  the self-intermediate scattering function of suspensions of hard spherical
  particles near the glass transition. \emph{Physical Review E} \textbf{1998},
  \emph{58}, 6073--6085\relax
\mciteBstWouldAddEndPuncttrue
\mciteSetBstMidEndSepPunct{\mcitedefaultmidpunct}
{\mcitedefaultendpunct}{\mcitedefaultseppunct}\relax
\EndOfBibitem
\bibitem[Kleban(1974)]{kleban_1974_toward}
Kleban,~P. Toward a microscopic basis for the de Gennes narrowing.
  \emph{Journal of Statistical Physics} \textbf{1974}, \emph{11},
  317--322\relax
\mciteBstWouldAddEndPuncttrue
\mciteSetBstMidEndSepPunct{\mcitedefaultmidpunct}
{\mcitedefaultendpunct}{\mcitedefaultseppunct}\relax
\EndOfBibitem
\bibitem[De~Gennes(1959)]{degennes_1959_liquid}
De~Gennes,~P. Liquid dynamics and inelastic scattering of neutrons.
  \emph{Physica} \textbf{1959}, \emph{25}, 825--839\relax
\mciteBstWouldAddEndPuncttrue
\mciteSetBstMidEndSepPunct{\mcitedefaultmidpunct}
{\mcitedefaultendpunct}{\mcitedefaultseppunct}\relax
\EndOfBibitem
\bibitem[Saito \latin{et~al.}(2016)Saito, Kobayashi, Masuda, Kurokuzu, Kitao,
  Yoda, and Seto]{saito2016slow}
Saito,~M.; Kobayashi,~Y.; Masuda,~R.; Kurokuzu,~M.; Kitao,~S.; Yoda,~Y.;
  Seto,~M. Slow dynamics in glycerol: collective de Gennes narrowing and
  independent angstrom motion. \emph{Hyperfine Interactions} \textbf{2016},
  \emph{237}, 1--8\relax
\mciteBstWouldAddEndPuncttrue
\mciteSetBstMidEndSepPunct{\mcitedefaultmidpunct}
{\mcitedefaultendpunct}{\mcitedefaultseppunct}\relax
\EndOfBibitem
\bibitem[Sobolev(2016)]{sobolev_2016_de}
Sobolev,~O. De Gennes Narrowing and Hard-Sphere Approach. \emph{The Journal of
  Physical Chemistry B} \textbf{2016}, \emph{120}, 9969--9977\relax
\mciteBstWouldAddEndPuncttrue
\mciteSetBstMidEndSepPunct{\mcitedefaultmidpunct}
{\mcitedefaultendpunct}{\mcitedefaultseppunct}\relax
\EndOfBibitem
\bibitem[Wu \latin{et~al.}(2018)Wu, Iwashita, and Egami]{wu_2018_atomic}
Wu,~B.; Iwashita,~T.; Egami,~T. Atomic Dynamics in Simple Liquid: de Gennes
  Narrowing Revisited. \emph{Physical Review Letters} \textbf{2018},
  \emph{120}, 135502\relax
\mciteBstWouldAddEndPuncttrue
\mciteSetBstMidEndSepPunct{\mcitedefaultmidpunct}
{\mcitedefaultendpunct}{\mcitedefaultseppunct}\relax
\EndOfBibitem
\bibitem[Mei \latin{et~al.}(2020)Mei, Zhou, and
  Schweizer]{mei2020thermodynamics}
Mei,~B.; Zhou,~Y.; Schweizer,~K.~S. Thermodynamics--structure--dynamics
  correlations and nonuniversal effects in the elastically collective activated
  hopping theory of glass-forming liquids. \emph{The Journal of Physical
  Chemistry B} \textbf{2020}, \emph{124}, 6121--6131\relax
\mciteBstWouldAddEndPuncttrue
\mciteSetBstMidEndSepPunct{\mcitedefaultmidpunct}
{\mcitedefaultendpunct}{\mcitedefaultseppunct}\relax
\EndOfBibitem
\bibitem[Mei \latin{et~al.}(2021)Mei, Zhou, and Schweizer]{mei2021experimental}
Mei,~B.; Zhou,~Y.; Schweizer,~K.~S. Experimental test of a predicted
  dynamics--structure--thermodynamics connection in molecularly complex
  glass-forming liquids. \emph{Proceedings of the National Academy of Sciences}
  \textbf{2021}, \emph{118}\relax
\mciteBstWouldAddEndPuncttrue
\mciteSetBstMidEndSepPunct{\mcitedefaultmidpunct}
{\mcitedefaultendpunct}{\mcitedefaultseppunct}\relax
\EndOfBibitem
\bibitem[Mirigian and Schweizer(2015)Mirigian, and
  Schweizer]{mirigian_2015_dynamical}
Mirigian,~S.; Schweizer,~K.~S. Dynamical Theory of Segmental Relaxation and
  Emergent Elasticity in Supercooled Polymer Melts. \emph{Macromolecules}
  \textbf{2015}, \emph{48}, 1901--1913\relax
\mciteBstWouldAddEndPuncttrue
\mciteSetBstMidEndSepPunct{\mcitedefaultmidpunct}
{\mcitedefaultendpunct}{\mcitedefaultseppunct}\relax
\EndOfBibitem
\bibitem[Ghosh \latin{et~al.}(2019)Ghosh, Chaudhary, Kang, Braun, Ewoldt, and
  Schweizer]{ghosh_2019_linear}
Ghosh,~A.; Chaudhary,~G.; Kang,~J.~G.; Braun,~P.~V.; Ewoldt,~R.~H.;
  Schweizer,~K.~S. Linear and nonlinear rheology and structural relaxation in
  dense glassy and jammed soft repulsive pNIPAM microgel suspensions.
  \emph{Soft Matter} \textbf{2019}, \emph{15}, 1038–1052\relax
\mciteBstWouldAddEndPuncttrue
\mciteSetBstMidEndSepPunct{\mcitedefaultmidpunct}
{\mcitedefaultendpunct}{\mcitedefaultseppunct}\relax
\EndOfBibitem
\bibitem[Ghosh and Schweizer(2019)Ghosh, and Schweizer]{ghosh_2019_microscopic}
Ghosh,~A.; Schweizer,~K.~S. Microscopic theory of the influence of strong
  attractive forces on the activated dynamics of dense glass and gel forming
  fluids. \emph{The Journal of Chemical Physics} \textbf{2019}, \emph{151},
  244502\relax
\mciteBstWouldAddEndPuncttrue
\mciteSetBstMidEndSepPunct{\mcitedefaultmidpunct}
{\mcitedefaultendpunct}{\mcitedefaultseppunct}\relax
\EndOfBibitem
\bibitem[Ghosh and Schweizer(2020)Ghosh, and Schweizer]{ghosh_2020_the}
Ghosh,~A.; Schweizer,~K.~S. The role of collective elasticity on activated
  structural relaxation, yielding, and steady state flow in hard sphere fluids
  and colloidal suspensions under strong deformation. \emph{The Journal of
  Chemical Physics} \textbf{2020}, \emph{153}, 194502\relax
\mciteBstWouldAddEndPuncttrue
\mciteSetBstMidEndSepPunct{\mcitedefaultmidpunct}
{\mcitedefaultendpunct}{\mcitedefaultseppunct}\relax
\EndOfBibitem
\bibitem[Xie and Schweizer(2020)Xie, and Schweizer]{xie_2020_a}
Xie,~S.-J.; Schweizer,~K.~S. A collective elastic fluctuation mechanism for
  decoupling and stretched relaxation in glassy colloidal and molecular
  liquids. \emph{The Journal of Chemical Physics} \textbf{2020}, \emph{152},
  034502\relax
\mciteBstWouldAddEndPuncttrue
\mciteSetBstMidEndSepPunct{\mcitedefaultmidpunct}
{\mcitedefaultendpunct}{\mcitedefaultseppunct}\relax
\EndOfBibitem
\bibitem[Saltzman and Schweizer(2006)Saltzman, and
  Schweizer]{saltzman2006activated}
Saltzman,~E.~J.; Schweizer,~K.~S. Activated hopping and dynamical fluctuation
  effects in hard sphere suspensions and fluids. \emph{The Journal of chemical
  physics} \textbf{2006}, \emph{125}, 044509\relax
\mciteBstWouldAddEndPuncttrue
\mciteSetBstMidEndSepPunct{\mcitedefaultmidpunct}
{\mcitedefaultendpunct}{\mcitedefaultseppunct}\relax
\EndOfBibitem
\bibitem[Ghosh and Schweizer(2020)Ghosh, and Schweizer]{ghosh_2020_microscopic}
Ghosh,~A.; Schweizer,~K.~S. Microscopic theory of onset of decaging and
  bond-breaking activated dynamics in ultradense fluids with strong short-range
  attractions. \emph{Physical Review E} \textbf{2020}, \emph{101}, 060601\relax
\mciteBstWouldAddEndPuncttrue
\mciteSetBstMidEndSepPunct{\mcitedefaultmidpunct}
{\mcitedefaultendpunct}{\mcitedefaultseppunct}\relax
\EndOfBibitem
\end{mcitethebibliography}

\end{document}